\documentclass[pdflatex,sn-mathphys]{sn-jnl}% Math and Physical

\usepackage{color}

\usepackage{physics}

\jyear{2023}%
\begin{document}

\title[]{Quantum state manipulation and science of ultracold molecules}

%%=============================================================%%

\author[1]{\fnm{Tim} \sur{Langen}}\email{t.langen@physik.uni-stuttgart.de}
\affil[1]{\orgdiv{5. Physikalisches Institut and Center for Integrated Quantum Science and Technology (IQST)}, \orgname{Universit\"at Stuttgart}, \orgaddress{\street{Pfaffenwaldring~57}, \city{70569 Stuttgart}, \country{Germany}}}

\author[2]{\fnm{Giacomo} \sur{Valtolina}}\email{valtolina@fhi-berlin.mpg.de}
\affil[2]{\orgname{Fritz-Haber-Institut der Max-Planck-Gesellschaft}, \orgaddress{\street{Faradayweg~4-6}, \city{14195 Berlin}, \country{Germany}}}

\author[3]{\fnm{Dajun} \sur{Wang}}\email{djwang@cuhk.edu.hk}
\affil[3]{\orgdiv{Department of Physics}, \orgname{The Chinese University of Hong Kong}, \orgaddress{\street{Shatin}, \city{Hong Kong SAR}, \country{China}}}

\author[4]{\fnm{Jun} \sur{Ye}}\email{Ye@jila.colorado.edu}
\affil[4]{\orgdiv{JILA and Department of Physics}, \orgname{National Institute of Standards and Technology and University of Colorado}, \city{Boulder}, \postcode{80309}, \state{Colorado}, \country{USA}}

\abstract{An increasingly large variety of molecular species are being cooled down to low energies in recent years, and innovative ideas and powerful techniques continue to emerge to gain ever more precise control of molecular motion. In this brief review we focus our discussions on two widely employed cooling techniques that have brought molecular gases into the quantum regime: association of ultracold atomic gases into quantum gases of molecules and  direct laser cooling of molecules. These advances have brought into reality our capability to prepare and manipulate both internal and external states of molecules quantum mechanically, opening the field of cold molecules to a wide range of scientific explorations.}

\maketitle

\section{Introduction}\label{sec:intro}

Over the past two decades research on cold molecules has blossomed from a nascent and somewhat exotic niche field into a strong scientific current that expands the horizon and lays new foundations for the physical sciences~\cite{Carr2009,Baranov2012,Bohn2017}.  The scientific community is witnessing a remarkable transition from early aspirations to impactful scientific fruition and emergent technology.  Pioneering ideas of cooling molecules down to unexplored low energy regimes~\cite{DiRosa2004,Stuhl2008} have led the way to the more mature pursuit of science goal-driven molecular quantum state control~\cite{Leung2023}. Chemical interactions are being studied with much greater details including individual reaction pathways and resonances~\cite{Hu2019,Bjork2016}. Molecular complexity has become a feature to demonstrate sophisticated quantum control and explore emergent phenomena~\cite{Changala2019,Liu2023,Mitra2020,Park2023,Yang2022,Zhang2022a}. Ideas for manipulating molecules with external fields for long-range, anisotropic interactions have brought the reality of tunable many-body Hamiltonians for studying exotic quantum dynamics~\cite{Yan2013,Hazzard2014,Li2023,Christakis2023}. Molecules with extended coherence times are now setting more stringent limits and opening novel grounds for exploring fundamental symmetry and new physics beyond the Standard Model~\cite{ACME2018,JILA2022}. Increasingly precise control of complex molecules fits right into the emergent theme of quantum information science that builds on high-fidelity manipulation of microscopic quantum systems~\cite{Ni2018,Hughes2020,Holland2022,Bao2022}.   

With the central role that molecules play in the physical sciences, this progress in the field of cold molecules is bringing together scientists from many different disciplines. Particle physicists are interested in using molecules to search for evasive particles and fields. Condensed matter physicists are building quantum material models based on cold molecules. Chemists dream about gaining elementary understandings of the most fundamental reaction processes to enable designer chemistry. Quantum scientists are using molecules to construct quantum simulation and information processing platforms. With more complex molecules there comes more powerful technologies that are attractive to biomedical researchers~\cite{Liang2021}.  

Molecules bring us great diversity along with rich energy level structure, and their control and exploitation for scientific exploration thus demand a wide scope of methods and approaches. While it is easy to judge the success of molecular cooling by their temperatures, scientific vision and purpose should always be the foremost guiding principle when designing and deciding research directions on cold molecules.  

In this brief review, we present only a selected few approaches that are being effectively pursued for cooling, trapping and manipulation of molecules, and a few examples of scientific successes that build upon the advances of these tools. These efforts underpin our goal of precise control of molecular states for achieving understanding of emergent complexity in quantum materials, controlled chemical processes, and powerful new methods for precision measurement and quantum information science. The purpose of this review is to provide a common connection between varying approaches of cold molecules and their relevant scientific goals. 

\section{Preparation of ultracold molecules}\label{sec:preparation}

In this first section we discuss recent progress in the cooling and trapping of molecules. We will focus on two approaches: association of quantum gases of atoms into ultracold molecules and direct laser cooling of molecules. These approaches are 
leading the production of molecular samples with high phase space densities and full control over quantum mechanical degrees of freedom.

We note that there exists a further large variety of powerful techniques and strategies to control the external and internal degrees of freedom of molecules, which we only mention here briefly. These include buffer gas~\cite{Hutzler2012} and supersonic sources~\cite{Segev2017,Wu2018}, decelerators~\cite{Meerakker2012} and their combination with conservative traps~\cite{Reens2017,Segev2019}, merged~\cite{Henson2012} or crossed~\cite{Vogels2015} molecular beams, as well as cryofuges~\cite{Wu2017} and Sisyphus cooling of electrically trapped molecules~\cite{Zeppenfeld2012}. We also note that there is a whole vibrant field of trapped molecular ions that have been used for studies of chemical reaction dynamics, precision measurement, and quantum logic gates~\cite{Sinhal2022,Puri2019,Mohammadi2021,JILA2022}.

\subsection{Association of quantum gases of atoms}
\begin{figure}[tb]
    \centering
    \includegraphics[width=0.73\textwidth]{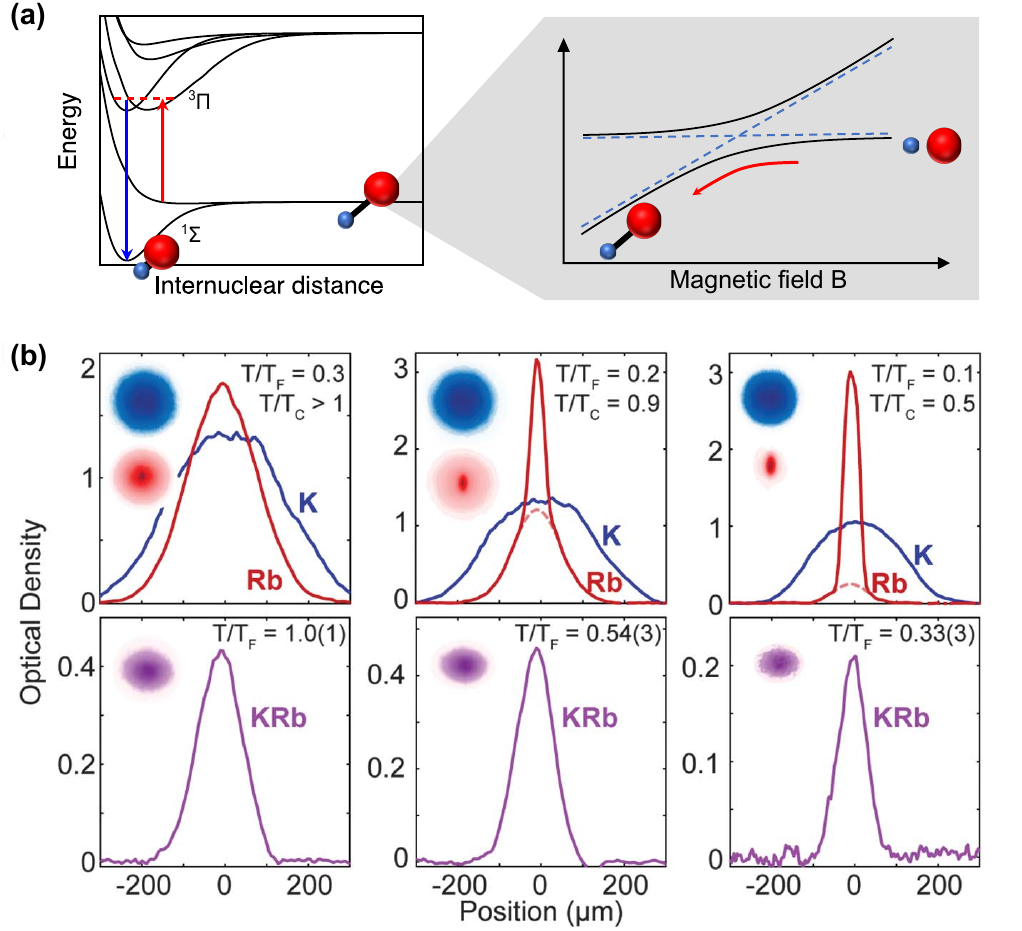}
    \caption{Preparation of ultracold molecules from ultracold atoms. (a) Magneto-association and \mbox{STIRAP} transfer. Pairs of atoms are associated into a weakly-bound molecule by sweeping a magnetic field over a scattering resonance (inset). These weakly-bound molecules are subsequently transfered into the rovibrational ground state by a two-color STIRAP via an excited state. Image from Ref.~\cite{Wang2015}. (b) Transition of a mixture of potassium and rubidium clouds (top) and associated KRb molecules from these clouds (bottom) into the quantum degenerate regime. Temperatures decrease from left to right. Image from Ref.~\cite{DeMarco2019}.}
    \label{fig:association}
\end{figure}

The first strategy to create ultracold molecules that we discuss here in detail, relies on the ability to create gases of ultracold atoms~\cite{Gadway2016,Moses2017}. By ramping an external magnetic field across a Feshbach resonance, pairs of these atoms can be associated into weakly bound, highly-excited Feshbach molecules~\cite{Kohler2006} (see Fig.~\ref{fig:association}a). Transfer of these molecules into their absolute ground state can be realized by stimulated Raman adiabatic passage (STIRAP). In this process, a pair of Raman lasers coherently transfers the molecules into their ground state via a suitable excited state. This removes thousands of Kelvins worth of binding energy from a gas at nanokelvin temperatures. The coherence of the process is essential for this, as any spontaneous processes would unavoidably lead to unacceptable heating. 

A key ingredient is the choice of the excited state which must accommodate the vastly different vibrational wavefunction extension as well as the different --- typically singlet and triplet --- characters between the Feshbach molecule and the ground-state molecule, respectively, to provide sufficient transition strengths for both legs of the Raman transfer. While such a state can in general be found through extensive spectroscopy with the help of detailed knowledge on the molecular structure, the Rabi frequencies, especially that for the upward transition, are often not very high. In this limit, high STIRAP efficiency can only be achieved with long-duration Raman pulses, which require high relative phase coherence between the Raman lasers. In earlier work, this was established by phase locking the Raman lasers to a stabilized frequency comb~\cite{Ni2008, Danzl2008}. Presently, the comb has been replaced by high-finesse ultrastable cavities with dual-wavelength coating~\cite{Aikawa2010}. With these efforts, one-way STIRAP transfer efficiency over 90\% has been achieved. The high efficiency is also important for detection, which relies on the reversed STIRAP process, to transfer ground-state molecules back to the Feshbach state.

Association of ultracold atoms has so far achieved the highest phase space densities in molecular gases. However, as this strategy requires ultracold atomic gases to start with, it only produces molecular species that are formed from atomic species that can be laser cooled. Typical experimental examples are heteronuclear~\cite{Ni2008,Takekoshi2014,Molony2014,Park2015,Guo2016,Rvachov2017,Voges2020,Stevenson2023} and homonuclear~\cite{Danzl2010} mixtures of alkalis, or mixtures of alkali and alkaline earth atoms~\cite{Barbe2018}. A related technique to magneto-association is photo-association of ultracold atoms, which has also been used to create bialkali molecules in the rovibrational ground state~\cite{Deiglmayr2008,Lang2008,Aikawa2010}. Further work has also suggested the possibility to form  alkali-lanthanide and alkaline earth-lanthanide mixtures with large spins~\cite{Frye2020,Schaefer2022}. 

Notably, recent progress has lead to the first generation of quantum degenerate molecular Fermi gases of KRb (see Fig.~\ref{fig:association}b) and NaK molecules~\cite{DeMarco2019,Valtolina2020,Schindewolf2022,Cao2023}. Moreover, the technique not only works for mesoscopic gases of many particles, but can be applied even on the single molecule level, where precisely two atoms can be turned into one molecule~\cite{Lui2018,Zhang2020,Ruttley2023}.

The molecular quantum gases created in this approach are readily loaded into optical dipole traps or optical lattices, leading to the first wave of studies of chemical reactions near absolute zero~\cite{Ospelkaus2010b}, long lifetimes of trapped molecules~\cite{Amodsen2012,Park2017}, tuning of state-dependent trapping potentials~\cite{Kotochigova2010,Neyenhuis2012} and spin-exchange interactions~\cite{Yan2013,Hazzard2014,Tobias2022,Christakis2023,Holland2022}.  

\subsection{Direct laser cooling}
Laser cooling relies on the repeated scattering of photons. Through this \textit{photon cycling}, the collective actions of tens of thousands of photons, each with a small individual momentum, lead to sizable forces for massive particles like atoms or molecules. Molecules with their complex internal structure thus do not, a priori, appear to be very well suited for laser cooling, as an excitation created by absorbing a photon can easily decay into many vibrational and rotational energy levels different from the initial one. As each of the corresponding transitions has a different frequency, photon cycling appears to require an unpractically large number of lasers to address all of these transitions. 

However, over the last decade it has been established that this challenge can be overcome in a large number of molecular species, by exploiting diagonal Franck-Condon factors~\cite{DiRosa2004} and transition selection rules~\cite{Stuhl2008} to limit vibrational and rotational branching. Ref.~\cite{Stuhl2008} further proposed practical schemes for constructing a molecular magneto-optical trap. While these requirements again limit cooling to a subset of molecular species, the required properties are fairly generic and provide a large variety of chemically diverse species~\cite{Tarbutt2018,McCarron2018review,Fitch2021}. 

\begin{figure}[tb]
    \centering
    \includegraphics[width=0.99\textwidth]{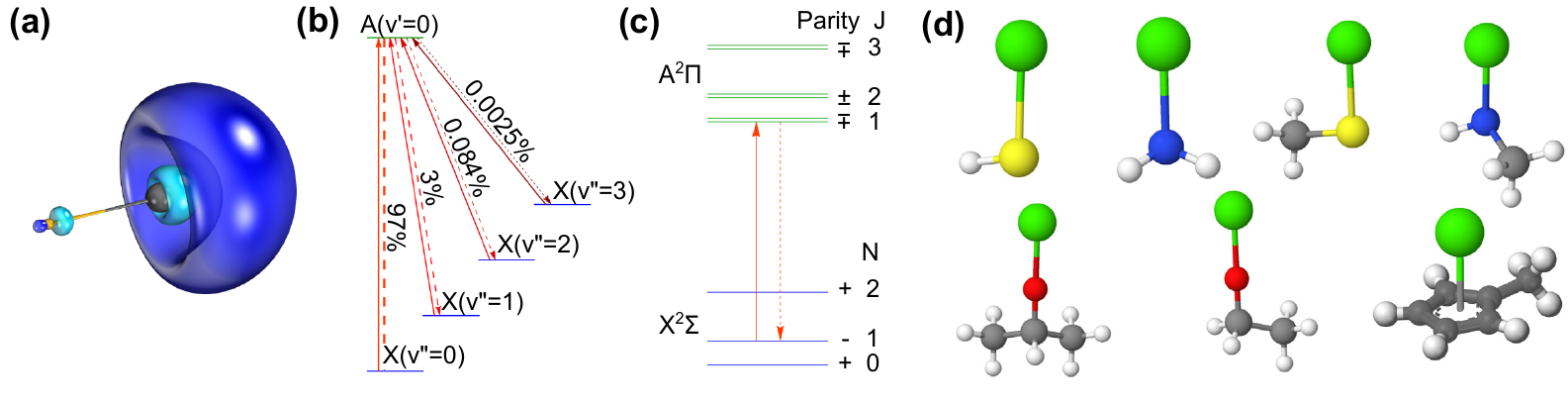}
    \caption{Laser cooling of molecules relies on a favorable molecular structure (a-c). The example in (a) shows the calculated valence electron distribution for the ground state of CaF molecules, where the optically active electron is centered around the metal atom and polarized away from the molecular bond. The molecule thus behaves similarly to an alkali atom, with transitions of the single valence electron independent from molecular vibrations to a very good approximation. Image from Ref.~\cite{AndereggPhD} (b) This leads to highly diagonal Franck-Condon factors, which strongly suppresses vibrational branching and limits the number of lasers required to realize a closed cooling cycle (c) In addition, rotational branching can be suppressed using parity selection rules, by driving transitions from the energetically lowest ground state with negative parity ($N=1$) to the lowest excited state with positive parity ($J=1,+$). Images from Ref.~\cite{Tarbutt2018}. (d) A similar strategy can be followed to laser cool also more complex polyatomic molecules that are formed from a suitable optically active cycling center (i.e. an atom or diatomic molecule, shown here in color) and a ligand (shown in grayscale). This ligand can, in principle, have very different levels of complexity, without significantly perturbing the laser cooling cycle. Image from~\cite{Augenbraun2020}.\\}
    \label{fig:lasercooling}
\end{figure}

The best understood examples exhibiting favorable vibrational properties are alkaline earth monofluorides (SrF~\cite{Barry2014}, CaF~\cite{Truppe2017,Anderegg2017}) and oxides like YO~\cite{Hummon2013,Yeo2015}. The calculated valence electron distribution of the CaF ground state is shown in Fig.~\ref{fig:lasercooling}a. While one of the calcium atom's two valence electrons forms the molecular bond with the electronegative fluorine atom, the other one is primarily located around the calcium atom and polarized away from the bond~\cite{Ellis2001}. Electronic excitations of this latter electron are thus nearly unaffected by changes in the molecular vibrational state, leading to diagonal Franck-Condon factors and strongly suppressed vibrational branching (Fig.~\ref{fig:lasercooling}b). The molecules thus behave --- in some ways --- similarly to alkali atoms that are common in atomic laser cooling experiments. In YO, the same principle holds~\cite{Stuhl2008} for two independent electron-nuclear spin angular momentum ground states, enabling efficient MOT and gray-molasses sub-Doppler cooling~\cite{Ding2020}. YO also features a narrow transition that can facilitate narrow line cooling similar to that in alkaline earth atoms~\cite{Collopy2015}.   

Many more species are under consideration or actively pursued: from light (MgF~\cite{Gu2022}) to heavy (YbF~\cite{Lim2018}, BaF~\cite{Chen2017,NLeEDM,Albrecht2020}, TlF~\cite{Norrgard2017}), very stable (AlF~\cite{Hofsaess2021}) to radioactive (RaF~\cite{GarciaRuiz2020}), as well as hydrides (BaH~\cite{McNally2020}, CaH~\cite{VazquezCarson2022}), chlorides (AlCl~\cite{Lewis2021}) or organic (CH~\cite{Schnaubelt2021}) species. Given the abundance of potential candidate molecules, this list is expected to further grow significantly in the coming years. 

In particular, these favorable vibrational properties are not limited to diatomics, but extend naturally to polyatomic species, where a large variety of ligands can be attached to atoms or diatomics acting as optical cycling centers~\cite{Isaev2016,Augenbraun2023} (Fig.~ \ref{fig:lasercooling}d). Initially being considered and demonstrated with linear triatomic molecules like SrOH, CaOH and YbOH~\cite{Kozyryev2017b,Vilas2022}, this approach has now also been extended experimentally to more complex molecules like $\mathrm{CaOCH}_3$~\cite{Mitra2020}. The limits of this approach are currently very actively explored on the theoretical chemistry side, with the goal of systematically predicting which ligands are compatible with which cycling centers~\cite{Augenbraun2023,Ivanov2020,Zhu2022}. 

On the practical side, the additional structure of polyatomics leads to additional vibrational bending, stretching and hybrid modes, which require additional lasers to address the corresponding states. While increasing the complexity of maintaining a nearly-closed optical cycle, this overall structure of polyatomics is particularly favorable for precision measurements~\cite{Kozyryev2017,Augenbraun2020,Maison2021} as it leads, e.g., to opposite parity, nearly degenerate doublets of states. Such molecules can thus generically be polarized in electric fields that are much smaller than the fields required for diatomic molecules and their structure provides a powerful tool to control systematic effects in precision measurements. Other applications of polyatomic molecules that have been considered include quantum simulation, quantum computation and ultracold chemistry and collisions~\cite{Augenbraun2023}.

For both diatomics and polyatomics, once vibrational branching is suppressed, angular momentum and parity selection rules for rotational states can be used in combination with remixing of dark states that are not addressed by the laser light, to realize a closed optical cycle~\cite{Stuhl2008,Yeo2015}(Fig.~\ref{fig:lasercooling}c). Recently, also the work on species with more complex hyperfine structure, which exhibit many additional dark states, has gained traction~\cite{Kogel2021,Zeng2023}. 

Turning a closed optical cycle into actual laser cooling requires the understanding of the resulting multi-level systems~\cite{Tarbutt2015,Devlin2016,Kogel2021}. Scattering rates in such multi-level systems are  proportional to $N_e/(N_g+N_e)$, where $N_e$ and $N_g$ are the number of exited and ground states that are coherently involved in the optical cycle. Typical monofluorides, for example, exhibit $4$ excited and $12$ ground states in each addressed vibrational level. The scattering rates are thus significantly reduced compared to the textbook two-level situation, which also goes hand in hand with increased saturation intensities for the transitions. Cooling of molecules therefore typically not only requires more lasers than cooling of atoms to address all required states, but also higher laser powers to achieve sufficiently fast scattering. 

However, apart from an increased level of complexity, one finds cooling forces similar to the ones known from atomic laser cooling, such as e.g. Doppler and Sisyphus forces. Molasses techniques can be used to image molecules down to the level of individual molecules~\cite{Cheuk2018}. Interestingly, and in contrast to atoms, closed optical cycles can also be realized in ways where absorption and emission occur on different wavelengths, which allows near background-free detection of molecules, e.g. for precision measurement applications~\cite{Shaw2021,Rockenhaeuser2023}. 

Experiments typically start with a slow molecular beam, formed e.g. by laser ablation and buffer gas cooling~\cite{Hutzler2012,Truppe2018}, which is laser-slowed~\cite{Barry2012,Yeo2015,Truppe2017slowing} to the capture velocity of a magneto-optical trap~\cite{Barry2014,Norrgard2016,Anderegg2017,Collopy2018,Vilas2022}. Further molasses cooling~\cite{Truppe2017,Ding2020,Vilas2022} then produces molecular samples that can be trapped in conservative potentials~\cite{McCarron2018,Anderegg2018,Williams2018,Wu2021,Langin2021}. 

The phase space densities that have been achieved in this way have increased by more than $10$ orders of magnitude in recent years, reaching now $10^{-6}$~\cite{Wu2021,Burau2022}, which is close to being suitable for further collisional cooling of the molecules to quantum degeneracy. Further progress is expected in the near future through more efficient slowing methods~\cite{Fitch2016,Petzold2018,Augenbraun2021,Langin2023}, more efficient sources~\cite{Jadbabaie2020}, and sympathetic cooling with atoms~\cite{Son2020,Jurgilas2021}. The current phase space densities are also sufficiently high to load optical tweezer arrays with single molecules~\cite{Anderegg2019}, enabling studies of collisions on the single molecule level~\cite{Cheuk2020}. 

%\newpage
\section{State engineering and coherence of single molecules}\label{sec:singlemolecules}

The rich internal structures of ultracold polar molecules, including their vibrational, rotational, and hyperfine states, have long been identified as great asset for many possible applications. A prerequisite for exploring this potential is the capability of controlling the internal states of the sample on the single quantum level. Building on the  progress in cooling discussed in the previous section, this has been fully achieved with the combined power of optical and microwave methods. In addition, microwave coupling between rotational levels has emerged as a major method for engineering and probing inter-molecular interactions. To fully explore the potential of this method, significant efforts have been devoted to understanding and controlling rotational coherence.
\newpage
\subsection{Internal state control}

In magneto-association, the weakly-bound Feshbach molecule is produced with a Feshbach resonance between atom pairs in single hyperfine Zeeman levels and is thus inherently in a single quantum level~\cite{Kohler2006}. For bi-alkali molecules, the ground electronic state is $^1\Sigma^+$, which has no electronic orbital and spin angular momenta. Their hyperfine structures are thus purely from the atomic nuclear spins and rotation~\cite{Aldegunde2008}. In magnetic fields, these structures split into $(2N+1)(2I_1+1)(2I_2+1)$ closely spaced levels, with $N$ the rotational quantum number, and $I_i$ the nuclear spin of the two atoms. Take $^{23}$Na$^{87}$Rb as an example, with $I_{\rm Na} = I_{\rm Rb} = 3/2$, there are 16, 48, and 80 levels for $N=0$, 1, and 2, respectively~\cite{Guo2018}. In a magnetic field, the total frequency span of the hyperfine levels in each rotational state is on the MHz level, while the intervals between adjacent hyperfine levels are even smaller. With the combination of selection rules and polarization, Rabi frequencies, and the Raman laser pulse lengths, population transfer to a single selected hyperfine levels has been realized routinely~\cite{Ni2008,Takekoshi2014,Molony2014,Park2015,Guo2016,Rvachov2017,Voges2020,Stevenson2023}. However, it is worth mentioning that high spectroscopic resolution and high population transfer efficiency have contradictory requirements on Rabi frequencies. This highlights again the importance of establishing good phase coherence between the Raman lasers, which will allow high STIRAP efficiencies at lower Rabi frequencies and longer pulses. If possible, using Feshbach resonances at higher magnetic fields to make hyperfine splitting larger also helps.

While most experiments start from molecules in the $v=0, N=0$ rovibrational level, STIRAP can also place the population in $N = 2$ directly. The $N = 1$ level, which cannot be reached by STIRAP as limited by parity selection rules, can be populated with a microwave driving the rotational transition, e.g. between $N=0$ and 1, as illustrated in Fig.~\ref{fig:onebodycontrol}a and b~\cite{Ospelkaus2010a,Will2016,Gregory2016,Guo2018}. In addition, two-photon microwave transitions are also frequently used for hyperfine manipulation. As has been demonstrated in RbCs, multiple microwave frequencies can also be applied to produce molecules in higher rotational levels~\cite{Blackmore2020b}. STIRAP can also transfer molecules to low lying excited vibrational states, such as $v = 1$~\cite{Guo2018}. For molecules with no two-body chemical reactivity in the $v=0$ level, this can serve as a knob to turn on the reaction~\cite{Ye2018}. 

For directly laser cooled $^2\Sigma^+$ molecules, after the MOT stage the population is distributed in $N = 1$, the state used for the realization of the closed laser cooling cycle (Fig.~\ref{fig:lasercooling}c). This manifold contains multiple spin-rotation and hyperfine levels. In recent experiments with CaF and SrF molecules, preparing the molecules into a single hyperfine Zeeman level has been demonstrated using optical pumping and microwave driving~\cite{McCarron2018,Williams2018}.
In combination with optical tweezers such state-control has facilitated the study of state-dependent collisions~\cite{Cheuk2020}, and entanglement of pairs of molecules via spin-exchange interactions~\cite{Holland2022,Bao2022}.

\subsection{Single molecule coherence}

The microwave coupling between rotational levels with opposite parities can induce strong dipolar interactions between molecules. This leads to a rotational dipolar spin-exchange interaction which can be used for realizing a variety of spin models~\cite{Gorshkov2011,Wall2015a} and implementing the iSWAP two-qubit quantum gate~\cite{Ni2018,Hughes2020}. However, to make these applications possible, the single molecule rotational coherence time must be longer than the time scale of the interaction. 

\begin{figure}[tb]
    \centering
    \includegraphics[width=0.94\textwidth]{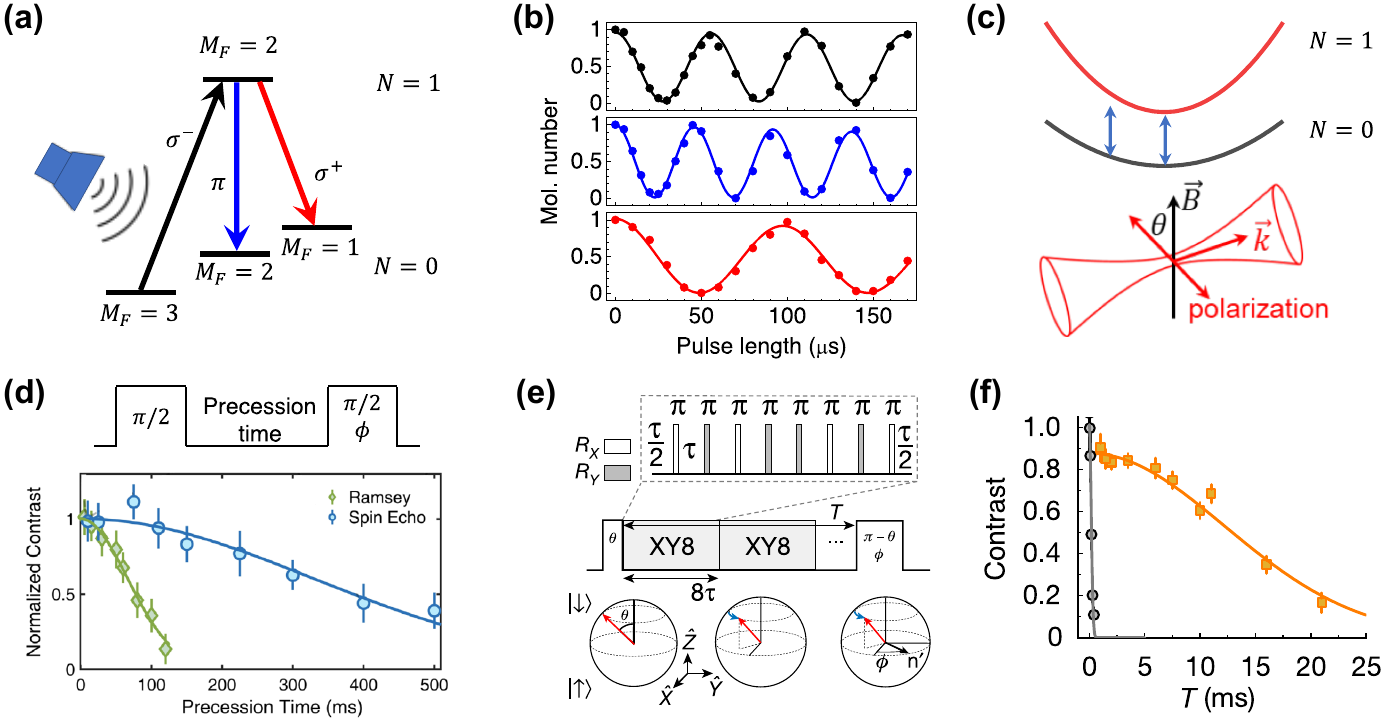}
    \caption{Internal state control and single molecule rotational coherence. (a) The schematic for the microwave driven transitions between rotational and hyperfine levels of NaRb. Here $M_F = m_N + m_i^{\rm Na} + m_i^{\rm Rb}$ is always a good quantum number, with $m_N$, $m_i^{\rm Na}$ and $m_i^{\rm Rb}$ the projections of the rotation and atomic nuclear spins. A nuclear spin flip is made possible by nuclear spin and rotation mixing. (b) Coherent population transfers from $N = 0$ to $N = 1$ or between hyperfine levels in $N = 0$. The oscillations and their corresponding transitions in (a) are color coded. (c) Differential light shifts cause rotational decoherence (top). This can be mitigated by tuning the polarization angle $\theta$ of the trapping light to the magic angle (bottom). (d) The rotational coherence time for single CaF molecules in optical tweezers is limited by a residual differential light shift (green diamond). With spin echo, the coherence time is extended to nearly $500\,$ms. Image from Ref.~\cite{Burchesky2021}. (e) Dynamical decoupling with the XY8 microwave pulse sequences is (f) powerful enough to extend the rotational coherence time of KRb molecules in 2D potentials to nearly 20 ms without the magic angle compensation. Images from~\cite{Li2023}.}
    \label{fig:onebodycontrol}
\end{figure}

For optically trapped molecules, the most important source of decoherence is the differential AC Stark shift~\cite{Kotochigova2010,Neyenhuis2012} which causes shifts of the rotational transition frequencies across the trap. This problem can be mitigated using the anisotropic polarizability of $N = 1$. By setting the polarization of the trapping light to the ``magic'' angle as depicted in Fig.~\ref{fig:onebodycontrol}c, the light shifts of $N=0$ and $N=1$ from the scalar polarizabilities can be made exactly the same. However, the polarizability of $N = 1$ also has a tensor term which mixes the different $m_N$ levels and results in a hyperpolarizability and a quadratic shift. This complication leads to an intensity dependent magic angle, and more importantly an incomplete differential light shift cancellation. The typical rotational coherence time, as measured by Ramsey spectroscopy, is limited to several ms for bulk samples in optical traps~\cite{Neyenhuis2012}. For single molecules in optical tweezers (Fig.~\ref{fig:onebodycontrol}d), a coherence time of 93(7) ms has been observed~\cite{Burchesky2021}.

Several different methods have been developed to make further improvements. Together with polarization angle adjustment, a moderate DC electric field can be applied to decouple nuclear spins from rotation to minimize the hyperpolarizability and thus the intensity dependent differential shift~\cite{Seeselberg2018,Blackmore2020}. Similarly, magnetic fields can be used to adjust the nuclear spin-rotation mixing to cancel the differential light shift at matched light intensities~\cite{Blackmore2020}. For NaRb in the ground band of the 2D optical lattices of a quantum microscope setup, a coherence time of 56(2) ms has recently been reported~\cite{Christakis2023}. A very different scheme is to create a ``magic'' trapping potential for the different rotational states by tuning the trapping light to near resonance with an excited molecular state. To minimize off-resonance scattering, the nominally forbidden $v=0  \leftrightarrow v=0$ transition between $X^1\Sigma^+$ and $b^3\Pi$ is used~\cite{Bause2020,He2021}. In RbCs, it is estimated that rotational coherence times greater than 1 s can be achieved in the magic potential~\cite{Cornish2023}.    

Imperfection of the magnetic field, both its fluctuations and gradients, is another important source of decoherence. For $^1\Sigma^+$ molecules, as the nuclear spins are very insensitive to magnetic fields, this is less of a problem. For $^2\Sigma^+$ molecules with a non-zero electron spin, the requirement on magnetic field is in general more stringent. However, the sensitivity to magnetic field can be reduced significantly by choosing a pair of rotational hyperfine Zeeman levels with small relative shifts so that the rotational transition frequency has only a quadratic Zeeman shift at low magnetic fields~\cite{Holland2022,Bao2022}.        

To extend the rotational coherence time further, more advanced microwave pulse sequences, such as spin echo and dynamical decoupling, can be used to remove decoherence from quasi-static sources. In several experiments, coherence times of hundred of milliseconds were achieved with these methods in both optical lattices~\cite{Yan2013,Christakis2023} and optical tweezers~\cite{Burchesky2021,Holland2022,Bao2022}. This is already much longer than the time scale of the dipolar interaction between molecules. Although more complex, dynamical decoupling pulse sequences are more powerful than spin echo~\cite{Holland2022,Bao2022}. In the latest KRb experiment~\cite{Li2023}, a coherence time of 17(1) milliseconds can still be achieved with an XY8 pulse sequence when implementing the magic angle is not realistic (Fig.~\ref{fig:onebodycontrol}e and f).     

Besides their insensitivity to magnetic fields, the nuclear spin hyperfine levels in $N = 0$ of $^1\Sigma^+$ molecules also experience very small differential light shifts~\cite{Park2017}. They can thus serve as storage qubits for preserving quantum coherence over long periods of time~\cite{Ni2018,Hughes2020,Tscherbul2023}. In several experiments, nuclear spin coherence times of several seconds have been observed in optical traps and lattices with two-photon microwave spectroscopy~\cite{Park2017,Gregory2021,Lin2022}. Although it has not been demonstrated, with spin echo or dynamical decoupling, the nuclear spin coherence time should be readily extended to over one minute.

\section{Molecular interactions}\label{sec:interactions}

The exquisite control of interactions in atoms has enabled collisional cooling of laser cooled atomic gases further into the ultracold regime, where collective quantum behavior plays a leading role. Such a gas of indistinguishable particles with finely tunable interactions performs as a quantum simulator, providing new insights in the behavior of complex many-body systems. Since the pioneering demonstration of the Mott insulator transition in an ultracold Bose gas \cite{Greiner2002} or the BEC-BCS crossover in Fermi gases~\cite{Randeria2014}, the field of quantum simulation with ultracold atoms has flourished \cite{Bloch2008}. Nonetheless, the class of phenomena that ultracold atoms can explore is restricted by the limited strength, range and isotropic nature of their interactions. 

Richer dipolar interactions with anisotropic and long-range character can be realized in magnetic quantum gases. This has recently led to the surprising observation of new states of matter like droplets and supersolids~\cite{Bottcher2021}. However, the strength of this interaction is still inherently set and limited by the atomic structure.  An ultracold and dense gas of polar molecules, with strong electric long-range interactions and a rich structure of coherent, long-lived states, is expected to greatly exceed the possibilities of such weakly dipolar atoms and realize never-observed-before phases of matter, such as topological superfluids \cite{Baranov2012,Schmidt2022}. 

However, the very same complexity that makes molecules so appealing renders the precise control of their interactions highly challenging. Molecular collisions at short distance are not as favorable as atomic ones: their dense spectrum of internal energy levels allows two colliding molecules to stick together~\cite{Mayle2013,Karman2019} and undergo a (photo-)chemical reaction that hampers collisional cooling and most of molecular quantum applications (Fig.~\ref{fig:collisions}a). Over the last decade, there has been tremendous efforts in understanding the collisional complexes that molecules form at short range, how they impact chemical reactions, and how we can suppress their formation with shielding methods~\cite{Avdeenkov2006,Gorshkov2008,Karman2018,Lassabliere2018}. Shielding has enabled the achievement of a new regime, a molecular gas dominated by elastic interactions, which can then be cooled to quantum degeneracy to study complex dipolar systems.

\subsection{Ultracold chemical reactions and shielding methods}

At ultralow temperatures, chemical reactions do not follow an Arrhenius-type equation since thermal energy is too low to overcome reaction barriers. Instead, wave-function overlap and quantum tunneling provide a way for molecules to meet at short distance and then react. Fermionic KRb ground-state molecules proved to be a valuable test bed for such ultracold chemistry. The reaction KRb + KRb $\rightarrow$ K$_2$ + Rb$_2$ is energetically allowed and proceeds according to a two-body loss rate. By changing the initial state of the molecular gas, it is possible to tune the reaction rate in a quantum-state-resolved fashion. Thanks to the centrifugal barrier in the \textit{p}-wave channel, a gas of identical fermionic molecules reacts less than a gas of distinguishable (or bosonic) molecules \cite{Ospelkaus2010b}, where collisions in the barrier-less \textit{s}-wave channel are allowed. Thus, control over the long-range barrier translates into control over the reactivity and stability of the molecules. This is the basic principle of reaction shielding. Since anisotropic dipolar interactions mix higher-order partial waves together, they provide a way to change the height and shape of the long-range barrier. 

\begin{figure}[tb]
    \centering
    \includegraphics[width=0.99\textwidth]{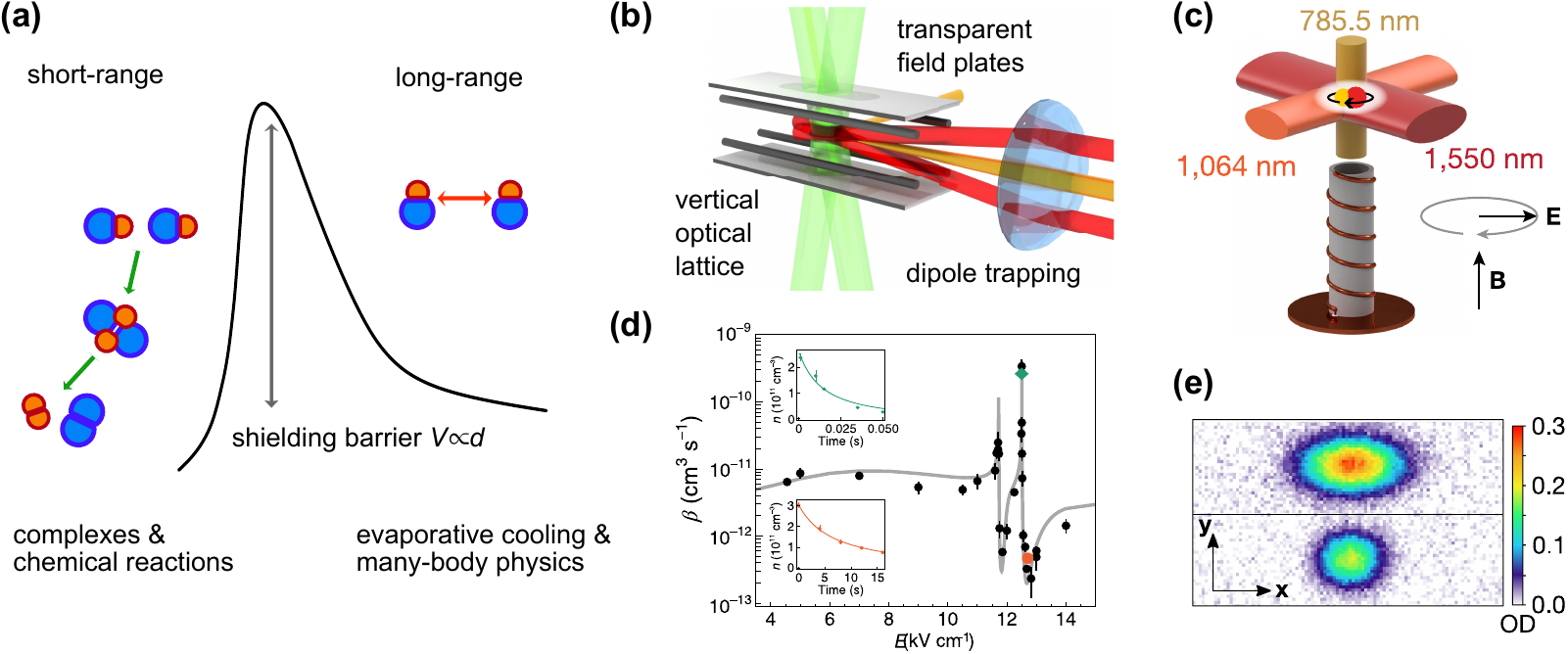}
    \caption{Controlling molecular interactions. (a) Short-range vs. long-range physics: The short-range regime happens inside the shielding barrier, whose height $V$ can be controlled via external fields. The long-range regime is instead outside the barrier. At short-range, molecular collisions are dominated by complex formation and chemical reactions, and often result in losses. Dipolar interactions enable elastic scattering beyond the barrier, which is crucial for evaporative cooling and many-body physics. (b) Experimental setup for collisional shielding in 2D gases of ultracold molecules. 2D trapping is realized by a combination of dipole traps, and a well-controlled DC electric field is applied by transparent bias field plates and auxiliary gradient rods. Image from Ref.~\cite{Valtolina2020} (c) Experimental setup for MW shielding of ultracold molecules. Circularly polarized MW electric fields are applied to molecules trapped in a combination of different dipole traps. Image from Ref.~\cite{Schindewolf2022} (d) Shielding resonance for KRb molecules in the $N=1$ state as a function of the bias DC electric field. The insets highlight the dramatic increase in molecular lifetime on different sides of the shielding resonance. Image taken from Ref.~\cite{Li2021} (e) Evaporative cooling of a 2D thermal Fermi gas of KRb molecules (top) to the quantum degenerate regime (bottom). Image taken from Ref.~\cite{Valtolina2020}.
    }
    \label{fig:collisions}
\end{figure}

Dipolar interactions can be induced by polarizing molecules in an external electric field. Initially, collisions of molecules in the lowest rotational level were investigated in a static electric field. While perpendicularly to the electric field dipolar interactions are repulsive, they are attractive along the field direction and facilitate short-range collisions. For a three-dimensional (3D) gas, the attractive dipolar component dominates and the rate of chemical reactions monotonically increases with the induced dipole moment \cite{Ni2010}. In two-dimensional (2D) traps, it is possible to retain only the repulsive side of dipolar interactions when the 2D gas is tightly confined along the field direction (Fig.~\ref{fig:collisions}b). This shielding method combines strong dipolar interactions with optical lattice confinement to exploit the stereodynamics of the reaction and suppress the reaction rate with respect to the 3D case \cite{deMiranda2011}. 

The technical complication in dealing with the large electric fields required for this type of shielding shifted the focus towards the creation of inherently chemically stable ultracold molecular samples \cite{Takekoshi2014,Molony2014,Park2015,Guo2016,Voges2020}. Surprisingly, the new species investigated for this purpose still displayed two-body losses close to the universal limit of chemically reactive molecules~\cite{Ye2018,Gregory2019,Guo2018b}. In 2019, a theoretical study \cite{Karman2019} analyzed the effect of optical dipole traps on molecular collisions, focusing on the role of intermediate complexes. Complexes are formed when molecules meet at short-range. For chemically stable molecules, energy conservation should allow complexes to dissociate back to the original reactants, while in the reactive case, complexes can transform into reaction products.  For the bialkali case, complexes have a broad optical absorption spectrum that peaks in the wavelength range of standard optical dipole traps. As a result, bialkali complexes quickly absorb trap photons, heat up, and are quickly lost from the trap. The detrimental role of optical trapping was later confirmed by two separate experiments \cite{Gregory2020,Liu2020}. However, the features of complexes are far from being completely understood, since additional experiments with different molecular species observed persisting losses even in absence of optical traps \cite{Bause2021, Gersema2021}.

Methods that shield molecular collisions from losses at short range completely are thus expected to become an integral part of molecular quantum gases experiments. 

Based on the the idea originally proposed in~\cite{Avdeenkov2006}, the first resonant shielding was achieved experimentally using molecules in large static electric fields, exploiting the rotational structure of polar molecules~\cite{Matsuda2020,Li2021}~(Fig.~\ref{fig:collisions}d).
In such a situation, resonant dipolar coupling results in a huge modulation of the chemical reaction rate in a narrow electric field region around the energy level crossing of pairs of rotational states coupled by the dipolar interactions. The shielding resonance enabled the realization of long-lived molecular gas in a static electric field. 

Similarly, based on other theoretical ideas~\cite{Karman2018,Lassabliere2018}, Ref.~\cite{Anderegg2021} exploited microwave  (MW) radiation to engineer an effective repulsion between two-body collisions of CaF molecules in optical tweezers. For proper detuning of the circularly polarized MWs, it was possible to suppress reaction rates by a factor of six even in tight tweezer traps.

Dressing with strong MW fields not only allows for shielding of collisions, but also shows great promise to independently control elastic interactions at short-range altogether. By finely tuning the ellipticity of the MW, it is possible to bring the open channel of two colliding molecules into resonance with a weakly-bound tetramer state~\cite{Lassabliere2018,Chen2023}. Control over the so-called field-linked resonance has enabled strong modulation of chemical reaction rates, but in the future it may be possible to coherently populate the tetramer state, similarly to the case of Feshbach molecules on a Feshbach resonance~\cite{Quemener2023}. Full control over resonant dipolar scattering and short-range interactions, for example by combining MW shielding with additional DC electric fields,  will enable the realization of the molecular analogue of the BEC-BCS crossover and the attainment and stabilization of molecular BECs~\cite{Schmidt2022}.

Resonances in molecular collisions can also be controlled with external magnetic fields. Similarly to ultracold atoms close to a Feshbach resonance, atom-molecule~\cite{Yang2019} and molecule-molecule ~\cite{Park2023} collisions can display resonant behavior, which results in a sharp increase of the loss rate of the trapped molecular gas. These resonances emerge when a long-range polyatomic (tri- or four-body) bound state can be brought to degeneracy with the scattering continuum. Precise control over their relative position enables new possibilities for coherent chemistry and molecule assembly. For instance, in collisions between sodium atoms and sodium–lithium molecules, an atom-molecule Feshbach resonance enabled the manipulation of the long-range barrier via control of the phase of the scattering wave function \cite{Son2022}. Interference between the long-range and the short-range part of the molecular potential resulted in a strong modulation of the reaction rate, even exceeding the universal limit.

\subsection{Evaporative cooling and quantum degeneracy}
Shielding kills two pigeons with one stone: while short-range losses are quenched from the effective shielding repulsion, elastic collisions at long-range are increased. Thus, shielding enables the realization of stable and strongly dipolar molecular gases, with large ratios of elastic-to-inelastic collisions. This condition was first demonstrated in 2D gases of fermionic KRb~\cite{Valtolina2020}, improving on the static shielding strategy of Ref.~\cite{deMiranda2011}. Cross-dimensional thermalization as a function of the induced dipole moment revealed an optimum spot around 0.2 D, where the separation among elastic and inelastic rate is maximum. This condition represented an ideal starting point for attempting evaporative cooling to quantum degeneracy, a necessary condition for any quantum simulation proposal. Controlled electric-field gradients enable to realize the first degenerate Fermi gas from an initial thermal distribution of the molecules, relying only on the strength of molecular interactions (Fig.~\ref{fig:collisions}e). The shielding resonance of Ref.\cite{Matsuda2020} was instead used to demonstrate evaporative cooling and phase-space-density increase in a 3D trap \cite{Li2021}.

Evaporative cooling to quantum degeneracy was also demonstrated for a MW-shielded gas of fermionic sodium-potassium (NaK) molecules\cite{Schindewolf2022} (Fig.~\ref{fig:collisions}c). The large effective dipole moment enabled by MW-shielding allowed to approach the hydrodynamic regime of collisions and elastic-to-inelastic collisions ratio of 500. Recently, MW-shielding lead to the stabilization of quantum gases of bosonic bialkali molecules, which could then be evaporatively cooled close to the onset of Bose degeneracy \cite{Bigagli2023, Lin2023}.

The newly discovered Feshbach resonances provide an alternative path for molecule formation close to quantum degeneracy. In Ref. \cite{Yang2022}, a new molecular quantum gas of weakly bound polyatomic molecules was demonstrated  by adiabatic magneto-association of potassium atoms and sodium-potassium molecules in the rovibrational ground state, which may even lead to the formation of ultracold ground state polyatomics at record-high phase space densities.

\begin{figure}[tb]
    \centering
    \includegraphics[width=0.78\textwidth]{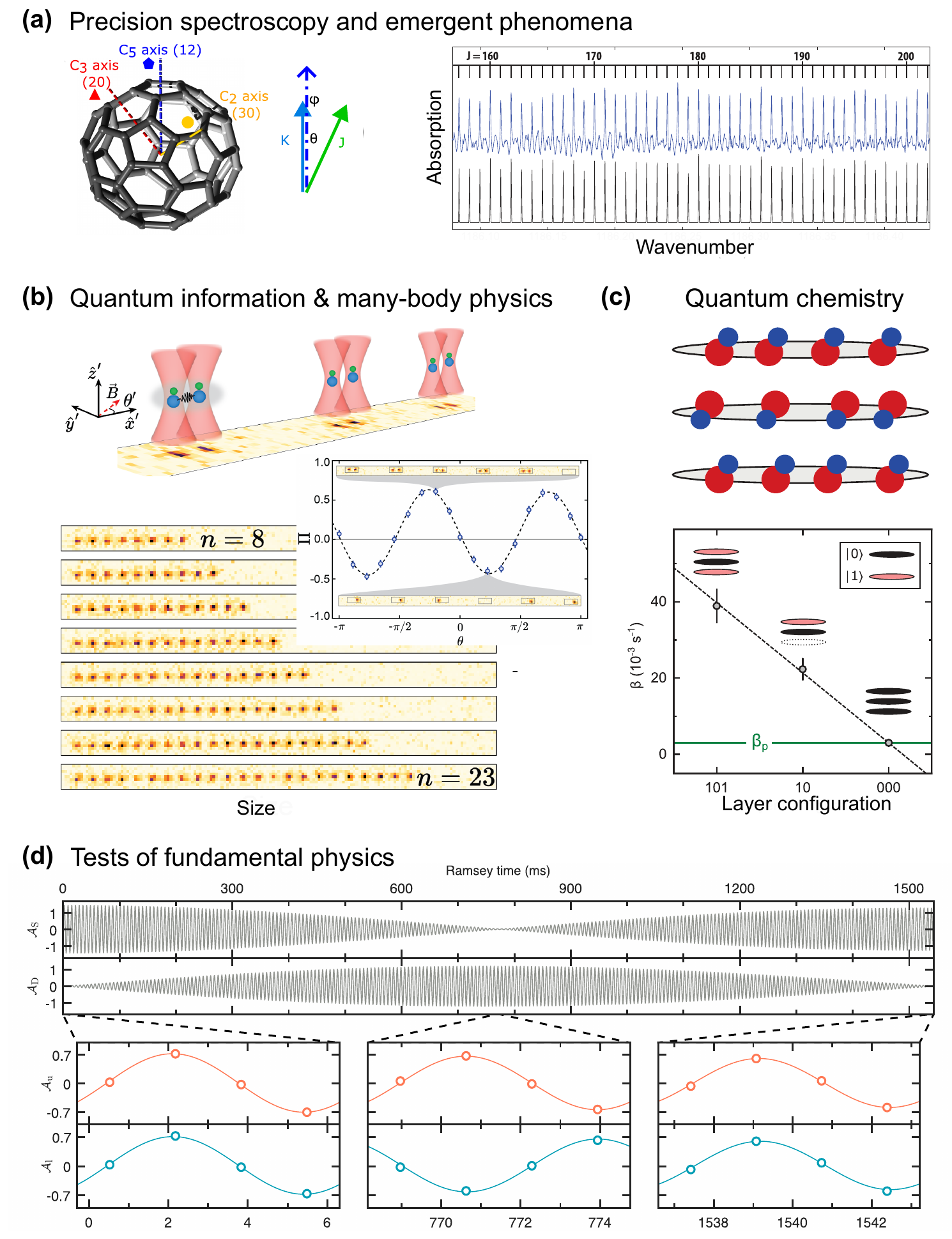}
    \caption{Advances in controlling intermediate-scale molecular systems. (a) Improved buffer gas cooling of large polyatomics, such as C$_{60}$, combined with advanced spectroscopy techniques, enables the probing of new features, such as ergodicity breaking in a mesoscopic quantum system. Figures from~\cite{Liu2023,Changala2019} (b) Tweezer arrays can be used to scale-up molecular systems from the single- to the many-molecule regime. Control over dipolar interactions in such arrays enables deterministic generation of an entangled Bell pair of individually trapped molecules \cite{Holland2022,Bao2022}. Figures from \cite{Holland2022}. (c) Tailored external fields can help engineer molecular structures with sub-wavelength resolution. This facilitates the tuning of chemical reactivity by stacking spatially separated two-dimensional layers of KRb molecules prepared in different rotational levels. Figures from~\cite{Tobias2022} (d) Ramsay interferometry with large, trapped samples of molecules showing coherence times exceeding a second, and providing the currently most stringent bound for the size of the electron's electric dipole moment. Figure from~\cite{JILA2022}.
    }
    \label{fig:outlook}
\end{figure}

\section{Scientific outlook}\label{sec:outlook}
Scientific explorations with cold molecules are now spreading across a variety of platforms for many scientific visions and goals, including precision measurement, cold chemistry, quantum simulation, and quantum information processing. Some of these topics are discussed in companion Reviews in this Issue. It is clear that we are still at the very early stage of realizing the full potential of the molecular quantum systems. Molecules are being employed in many-body physics experiments of increasing complexity (Fig. \ref{fig:outlook}b), but many exotic quantum phases and iconic quantum matter remain unexplored or out of reach with the current capabilities. NISQ-era quantum simulation employing interacting molecules offers fascinating opportunities to study exotic quantum phases and dynamics. Molecule-based precision measurements are setting new limits for violations of fundamental symmetries. When probed with state-of-the-art quantum control techniques, even molecular systems that have been known for decades reveal surprising, emergent phenomena (Fig.~\ref{fig:outlook}a). These achievements further support the dream of realizing 
fully quantum state-engineered molecules that are designed to have optimized sensitivity for fundamental physics. A first step in this direction is the trapping of large scale samples of suitable molecules, including also polyatomics with favorable level structure for precision measurement applications (Fig.~\ref{fig:outlook}d). Incorporating modern spectroscopy and detection tools with cold molecules are allowing us to follow reaction pathways and kinetics, and steer these reaction processes with external fields and unprecendented spatial resolution (Fig.~\ref{fig:outlook}c). The exciting prospect of using quantum information science to explore complex molecular structure and uncover hidden interaction dynamics will breathe powerful new life into quantum chemistry. Entanglement operations are being demonstrated on molecules trapped in optical fields, signaling initial steps in molecule-based quantum information processing (Fig. \ref{fig:outlook}b), but key challenges remain to achieve high fidelity and scalability.

\backmatter
\bmhead{Acknowledgments}
We thank P. Aggarwal, B. Augenbraun, L. Liu, and A. M. Rey for comments and suggestions on the manuscript. T.L. acknowledges support from Carl Zeiss Foundation, the RiSC programme of the Ministry of Science, Research and Arts Baden-W\"urttemberg, and the European Research Council (ERC) under the European Union’s Horizon 2020 research and innovation programme (Grant agreement No. 949431). G.V. acknowledges support from the Alexander von Humboldt Foundation. D.W. is supported by Hong Kong RGC General Research Fund (Grants No. 14301818 and No. 14301119) and Collaborative Research Fund (Grants No. C6009-20GF). J.Y. acknowledges support from ARO and AFOSR MRUI, and NIST. 

\section*{Declarations}

\paragraph{Author contributions.} All authors contributed to the writing of the manuscript.

\paragraph{Author information.} The authors declare no competing financial interests. Correspondence and request for materials should be sent to T.L. (t.langen@physik.uni-stuttgart.de), G.V. (valtolina@fhi-berlin.mpg.de), D.W. (djwang@cuhk.edu.hk), and J.Y. (ye@jila.colorado.edu). 

\newpage
\bibliography{bibliography}% common bib file

\end{document}